\documentclass[%
 reprint,
superscriptaddress,
 amsmath,amssymb,
 aps,
]{revtex4-2}

\usepackage{graphicx}
\usepackage{dcolumn}
\usepackage{bm}

\begin{document}


\title{Physics-Informed Machine Learning for Optical Modes in Composites}

\author{Abantika Ghosh}
\affiliation{Department of Physics and Applied Physics, University of Massachusetts Lowell, Lowell, MA 01854, USA}

\author{Mohannad Elhamod}
\author{Jie Bu}
\affiliation{Department of Computer Science, Virginia Tech, Blacksburg, VA 24060, USA}

\author{Wei-Cheng Lee}
\affiliation{Department of Physics, Applied Physics, and Astronomy, Binghamton University - State University of New York, Binghamton, 13902, USA}

\author{Anuj Karpatne}
\affiliation{Department of Computer Science, Virginia Tech, Blacksburg, VA 24060, USA}

\author{Viktor A Podolskiy}
\affiliation{Department of Physics and Applied Physics, University of Massachusetts Lowell, Lowell, MA 01854, USA}
\email{\authormark{*}viktor\_podolskiy@uml.edu} 

\date{\today}

\begin{abstract}
We demonstrate that embedding physics-driven constraints into machine learning process can dramatically improve accuracy and generalizability of the resulting model. Physics-informed learning is illustrated on the example of analysis of optical modes propagating through a spatially periodic composite. The approach presented can be readily utilized in other situations  mapped onto an eigenvalue problem, a known bottleneck of computational electrodynamics. Physics-informed learning can be used to improve machine-learning-driven design,  optimization, and characterization, in particular in situations where exact solutions are scarce or are slow to come up with. 
\end{abstract}

\maketitle

\section{Introduction}

Optical composites (metamaterials and metasurfaces) emerge as  flexible platforms for novel optical applications that include superimaging, planar lensing, and sensing \cite{hasman1,capassoMetasurface,shalaevMetasurface,metasurfaceYu,metasurfaceCapasso2,metasurfaceTsai,metasurfaceBrongersma,eleftheriadesMetasurface,PCBook, ShalaevBook, vpBook}. In these applications, computationally expensive design and optimization of the structure of optical composites represents a crucial bottleneck. Machine learning (ML)  has been instrumental in addressing some needs of the photonics community\cite{MLPhoto.1,MLPhoto.2,MLPhoto.3,MLFan,MLSuchowski,MLCai,MLLiu,MLLiu2,MLFan2,MLFainman,MLImagingGhosh,MMMLFan} to the point that ML is sometimes predicted to overtake the scientific development process itself\cite{endOfTheory}. However, conventional ML algorithms often require large data-sets to produce properly trained, generalizable models\cite{MLbook}. Therefore, ML deployment for optimizing complex composites is often slow and problematic. Several approaches that mitigate the required size of the training set, for example by training in parameter sub-space  that minimizes the uncertainty of the resulting model\cite{smartLearning,soljacic} have been recently proposed. Attempts to build Physics-Informed ML, incorporating analytical equations into ML learning process itself have shown promise in simple differential equations, as well as in physics of fluids and in imaging\cite{PGintro,PGturbulence,PGMLJi}. Here, we present physics-informed ML  for optical composites, and illustrate the proposed formalism on the example of solving for the modes of a composite with periodic permittivity profile, achieving fast and highly generalizable predictions with relatively small training datasets. 


We develop the class of ML models that map the spatial profile of permittivity of the composite to the combination of the propagation constant and parameters that determine spatial behavior of the mode supported by the system (Fig.\ref{composite}). Specifically, the developed ML process predicts the properties of highest-effective index TM-polarized mode propagating in a multi-layer periodic composite whose unit cell contains 10 layers. Several sets of composites, some purely dielectric, and some plasmonic are used to assess accuracy and generalizability of the resulting models. The resulting ML models fully utilize the benefits of parallelism offered by Graphics-Processing-Unit (GPU) computing that are unavailable to iterative eigenvalue solvers\cite{linalg}.


Note that although we present data for 1D composites, the mathematical formalism used to map Maxwell equations to an eigenvalue problem, rigorous coupled wave analysis (RCWA)\cite{RCWA,normalVector}, can be directly used for periodic and non-periodic 2D media\cite{non-periodic1,non-periodic2,non-periodic3}.

\begin{figure*}
    \centering
         \includegraphics[width=0.9\textwidth]{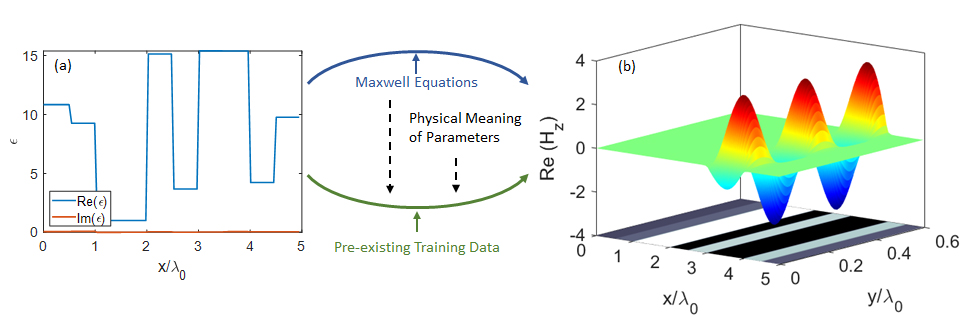}
        \label{composite}
     \caption{(a) An example of the distribution of dielectric permittivity across one period of the multilayered composite and (b) spatial dependence of the highest-index propagating mode supported by this composite. Solid arrows illustrate Physics- and ML-based approaches to solving Maxwell equations, with dashed line illustrating the new Physics-Informed ML described in this work.}
     \label{composite}
\end{figure*}

\section{Exact solutions of Maxwell equations}

In the case of periodic layered materials considered in this work (with $x$ being the direction of layer growth), solutions to Maxwell equations can represented as a linear combination of modes with either transverse-electric (TE, $E_z\neq 0, H_z=0$) or transverse-magnetic (TM, $E_z=0, H_z \neq 0$) polarization, with full Maxwell equations reducing to independent partial differential equations for $E_z$ and $H_z$ components, respectively, and with  remaining ($x,y$) components of the fields expressed in terms of their $z$ counterparts. In particular, propagation of the TM-polarized waves is given by: 
 \begin{equation} \label{eq:1}
     \frac{\partial^2{H_z} }{\partial{y^2}}=-\epsilon(x)\left[\frac{\partial{}}{\partial{x}}\left(\frac{1}{\epsilon(x)}\frac{\partial}{\partial{x}}\right)+\left(\frac{\omega}{c}\right)^2 \right] H_z, 
 \end{equation}
with $\omega$ being angular frequency and $c$ being speed of light in vacuum.  

RCWA\cite{RCWA,normalVector}, a semi-analytical method to analyze the mode structure of spatially periodic composites, takes explicit advantage of the periodicity and uses the Fourier expansion of spatial profile of both permittivity and electromagnetic fields: 
   \begin{eqnarray}\label{eq:2}
      \epsilon(x)&=&\sum_m{\varepsilon_m e^{-i q_m x}}\nonumber \\
      H_z(x,y)&=&e^{ik_y y}\sum_m{h_m e^{i k_{xm} x}}
  \end{eqnarray}
 to convert differential Eq.(\ref{eq:1}) to an eigenvalue problem: 
  \begin{equation}\label{eq:3}
     \sum_j\hat{A}_{m j} h_j=k_y^2 h_m
 \end{equation}
with 
\begin{equation}\label{eq:4}
\hat{A}_{m j} =-\sum_s{\varepsilon_{s-j}k_{xm}\varepsilon_{m-s}^{-1}k_{xs}}+\left(\frac{\omega}{c}\right)^2\varepsilon_{m-j}.    
\end{equation}
 In the equations above $q_m=\frac{2\pi m}{\Lambda}$ is the multiple of the reciprocal lattice constant, $\Lambda$ is the period of the composite, $k_{xm}=k_0+q_m$, $\varepsilon^{-1}_m$ represent the Fourier coefficients of $1/\epsilon(x)$, and parameter $k_0$ plays the role of the $x$ component of the quasi-wavenumber of the mode.

In all practical realizations, finite Fourier expansions (rather than Fourier series) have to be used. The complexity of the composite determines the number of terms in Fourier expansions that are required for adequate representation of permittivity and electromagnetic fields, and in turn determines the size of the matrix $\hat{A}$.
As the complexity increases, direct solution of Eq.\eqref{eq:3} becomes increasingly slow and resource-intensive, motivating the development of tools that avoid the direct solution of the eigenvalue problem, such as ML-assisted mode analysis presented here. 

To comprehensively assess the performance of the ML-based models we generated three datasets, representing geometry of the particular composite and the (highest-$k_y$) mode propagating in this composite. In these studies, the size of the unit cell was fixed at $\Lambda=5\lambda_0$  (with $\lambda_0=2\pi c/\omega$ being free-space wavelength) and each period of the composite was assumed to contain 10 layers of identical thickness, the quasi-wavenumber of the mode was parameterized by the angular parameter $\theta$ via $k_0=\frac{\omega}{c}\sin\theta$, and Fourier expansions contained the components corresponding to $m\in[-m_{\rm max},m_{\rm max}]$ with $m_{\rm max}=50$ in Eq.\eqref{eq:2}.   

The first two (photonic) sets contained data for lossy ($0<\epsilon''<0.1$) low-index $(1<\epsilon'<4)$ and high-index $(1<\epsilon'<16)$ dielectric stacks, with randomly-assigned permittivity for each sub-layer. The remaining set was similar to the high-index dielectric set with  25\% of configurations containing plasmonic sub-layers with permittivity of $\epsilon=-100+25i$. Each set contains data for 2000 geometries with $\theta\in[20^o,40^o,60^o,80^o]$ for each geometry. Overall, each dataset contained 8000 combinations, mapping the configuration, parameterized by 21 real numbers $\{\theta,\epsilon',\epsilon{''}\}$, onto a set of 204 real numbers that represent real and imaginary parts of $k_y$ and $h_m$ [see Appendix]. 

Fig.~\ref{nEff} illustrates the distribution of propagation constants of the modes within each dataset. Note that the propagation constants of the modes in the dielectric composites is constrained by the largest refraction index within the set. In contrast, plasmonic dataset contains the modes with very high propagation constants that originate as result of the interplay between different surface plasmon polaritons\cite{vpBook} 

\begin{figure}[htb]
    \centering
    \includegraphics[width=7cm]{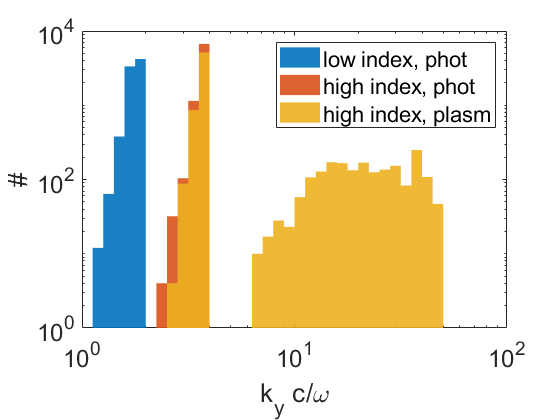}
     \caption{Distribution of the propagation constants of the modes within the three datasets used in this work; each dataset contains 8000 configurations}
     \label{nEff}
\end{figure}

\section{Black box- and Physics-informed ML}

Artificial Neural Networks (ANN) emerge as robust and flexible tools capable of deducing the input$\rightarrow$ output dependencies within the data\cite{MLbook}. The ANNs, inspired by biological neural networks, contain a set of linear coupling layers and a set of nonlinear activation layers stacked in-between the coupling layers. The typical ANN feeds the input into its first layer; information then flows within the ANN as output of one layer becomes the input for the subsequent layer; the output of the final layer represents the ML-based prediction. During the training stage, coupling coefficients that define the information flow are adjusted to minimize the {\it loss function}, deviation between the ANN-based prediction and the known exact results (ground truth). After training, the coupling coefficients are fixed and the ANN-based model is ready for deployment. 


In this work we use three different approaches to train the ANNs-based models. First, the default physics-agnostic ``black box'' formalism is used. In this approach, mean-squared deviation between components of predicted and ground truth sequences are used as optimization criterion during the training process\cite{MLbook}. In the second, meaning-informed approach, the loss function is adjusted to explicitly utilize the fact that the output sequence contains the combination of the eigenvalue and the components of the eigenvector. Meaning-informed loss is used in the final approach as well. In addition, this Physics-informed approach uses the input sequence to generate the matrices $\hat{A}$ within the network during the training process and to enforce the Eq.\eqref{eq:3} as additional constraint during the training, aiming to produce the explicitly physics-consistent results. Apart from the implementation of the loss function, the topology of the three ANNs is identical (see Appendix for more details). 

Our implementation of the physics-informed training follows the recipe for dynamic adjustment of the loss\cite{VT} that has been shown to improve the convergence of the model that uses multiple competing objectives\cite{MLhard}. Note however, that in contrast to Ref.\cite{VT}, the proposed formalism calculates the matrix $\hat{A}$ only during training but not during prediction stage. Moreover, even during training, the model does not directly solve the eigenvalue problem but rather tests the validity of the ML predictions against Eq.\eqref{eq:3}. A single matrix multiplication that is required for such testing is much faster than the iterative algorithms that often underline eigenvalue solvers\cite{linalg}. 

In addition to enforcing the consistency with Maxwell equations, Physics-informed training enables the expansion of the training set by ``padding'' it with configurations of composites that the model may expect to see in future deployments. As we show below, even in absence of full solutions, these configurations (termed {\it unlabeled data} below) can significantly improve the quality of the resulting ML model.  


\begin{figure}[tbh]
         \centering
         \includegraphics[width=7cm]{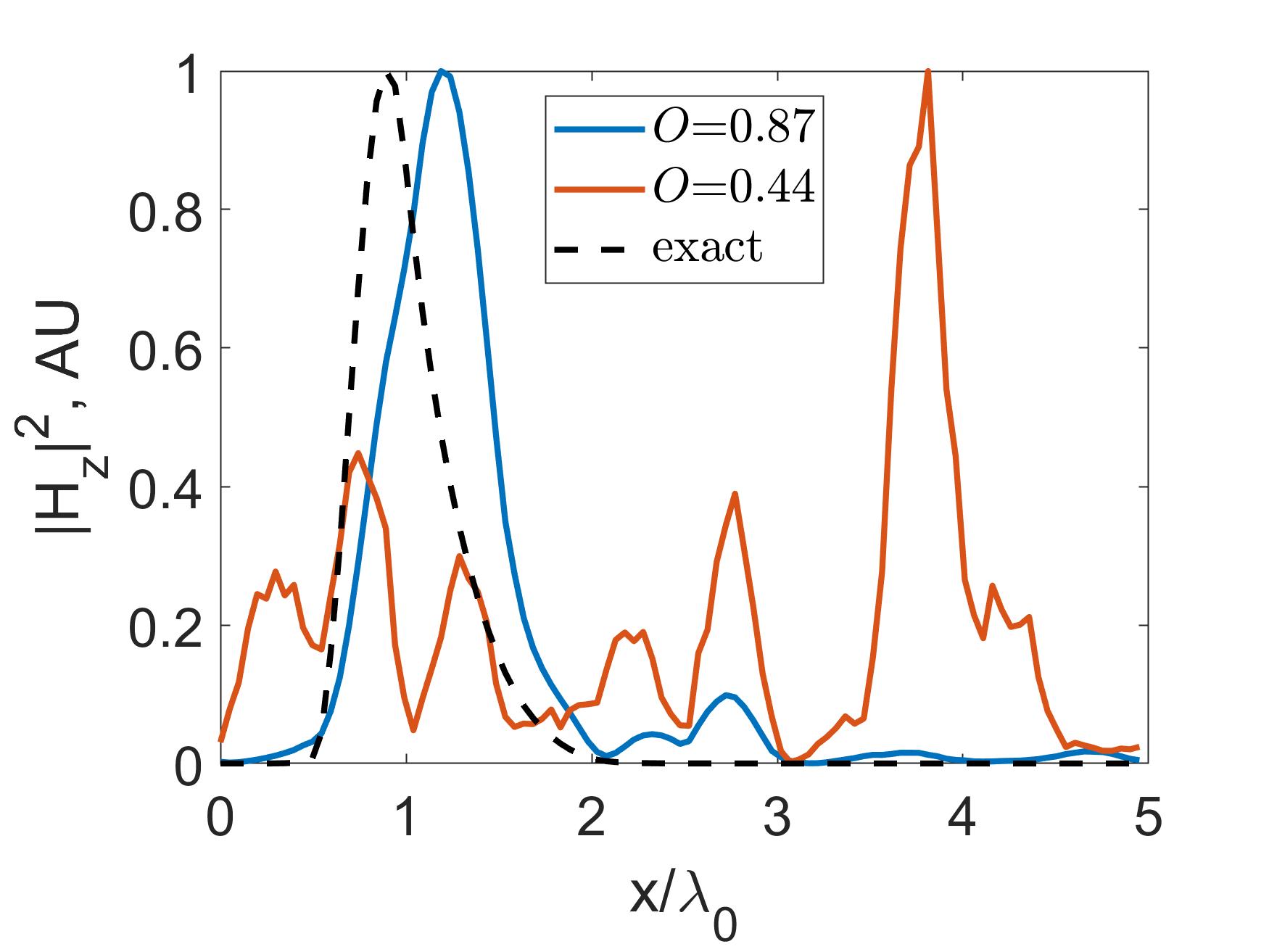}
         \caption{Illustration of the relationship between the numerical value of overlap parameter $O$ and the quality of prediction of spatial profile of the mode}
         \label{overlap}
 \end{figure}
 
A properly trained ML model should correctly predict the propagation constant of the mode as well as its spatial profile. We therefore use two (dimensionless) parameters to characterize the performance of the ML models, the normalized error in prediction of propagation constant $\delta$ and the modal overlap across the unit cell $O$ (see Appendix for derivation),
\begin{eqnarray} \label{eq:5}
\delta&=&\left|\frac{{k_y}-k_y^{\rm gt}}{k_y^{\rm gt}}\right|, \nonumber
\\
O&=&\frac{\left|{\sum}_m{h_m^*} h_m^{\rm gt}\right|}{\sqrt{\sum_m |h_m|^2}\sqrt{\sum_m |h_m^{\rm gt}|^2}}
\end{eqnarray}
In the expressions above, the quantities with "gt" superscript represent the ground truth, while the quantities with no superscript represent ML predictions and ``*'' corresponds to complex conjugation. 

Figure \ref{overlap} illustrates the relationship between the value of the parameter $O$ and the agreement between the predicted and exact profiles of magnetic field across the unit cell. It is seen that $O\gtrsim 0.8$ represent the adequate prediction quality.

 

\section{Results and Discussion}

\begin{figure}
    \centering
    \includegraphics[width=8cm]{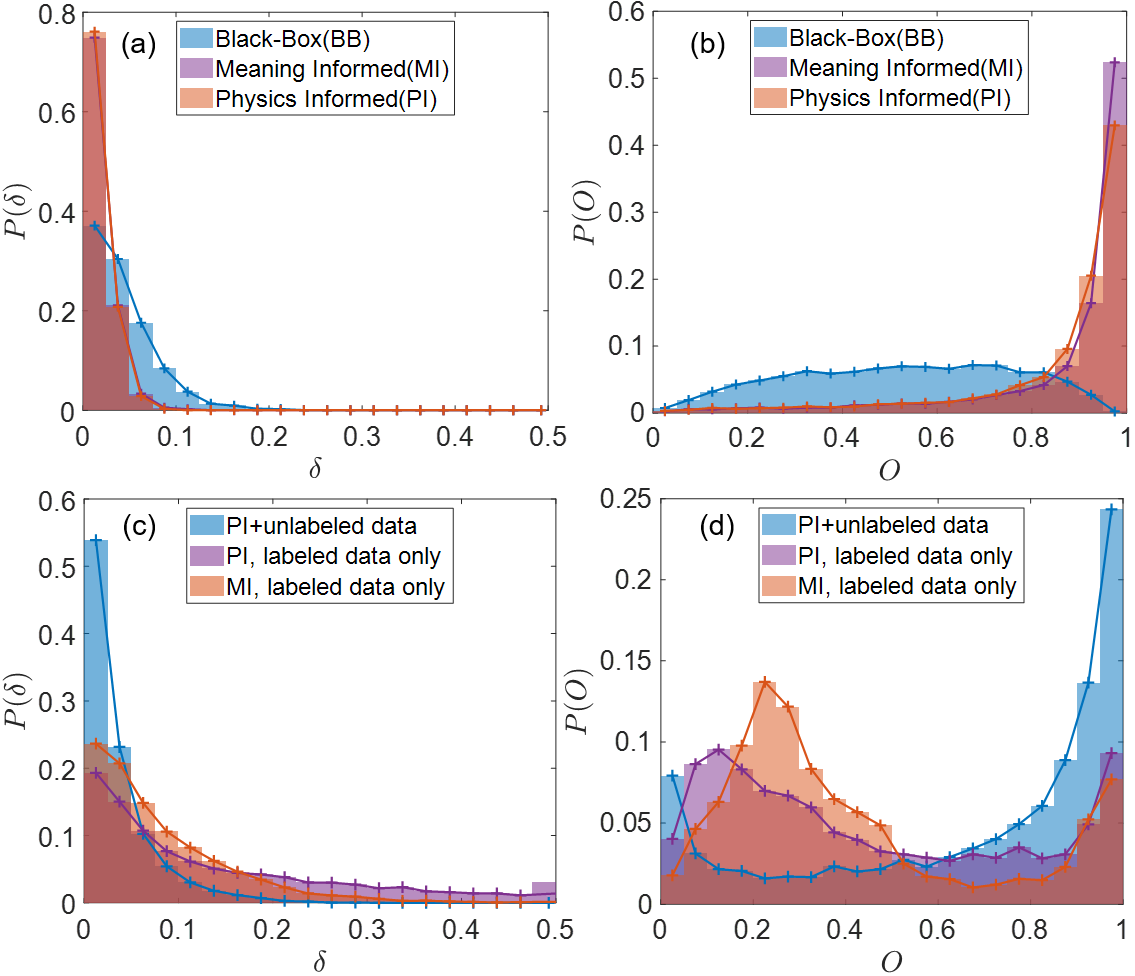}
    \caption{The accuracy of predicting the propagation constant (a,c) and spatial profile (b,d) of modes from the low-index photonic dataset by different ML models trained on $\sim$800 configurations/model (a,b); 
    and with models trained on a subset representing $\sim$200 configurations with pre-selected value of parameter $\theta$; additional configurations (without corresponding mode solutions) are used as unlabeled data for one of the physics-informed models}
    \label{diffModel}
\end{figure}

Figure \ref{diffModel} illustrates the typical performance of the different ML models that are trained on 10\% ($\sim$800 randomly selected configurations) of  one of the datasets, tested on predicting the modes of the remaining configurations within the same dataset. Note that both meaning- and physics-informed models drastically outperform their black-box counterpart. Further analysis (see Appendix) demonstrates that significant expansion of the training subset can improve the performance of the BB model. The requirement to have a large training set represents the main limitation of conventional science-agnostic ML, the limitation that makes such models virtually impractical in applications where the training sets are scarce (due to, for example, significant time that it takes to solve Maxwell equations in composites with complex geometries).

All models lose accuracy when the training set is further decreased by considering only $\sim$200 configurations, a subset of initial training pool representing one pre-selected value of parameter $\theta$ [see Eq.\eqref{eq:2}], Figure \ref{diffModel}(c,d). However, the performance of the Physics-informed model can be significantly improved by expanding its very limited training set with unlabeled configurations. Note that providing these extra configurations -- with no corresponding solutions -- brings the model performance almost in line with the baselines that have 4-times larger training libraries of propagation constants and mode profiles. 

One of the main limitations of ML models is their limited ability to predict the properties of systems that significantly deviate from their training data. As seen in Fig. \ref{nEff}, the datasets used in this work are designed to have significantly different distributions of optical modes. As result, the models trained on a particular dataset perform best on predicting the propagation constants and mode profiles for configurations within the same dataset. The models trained on plasmonic dataset also perform well on predicting properties of high-index photonic dataset (that essentially represents a subset of its plasmonic counterpart; see Appendix). 

\begin{figure}
\centering
\includegraphics[width=8cm]{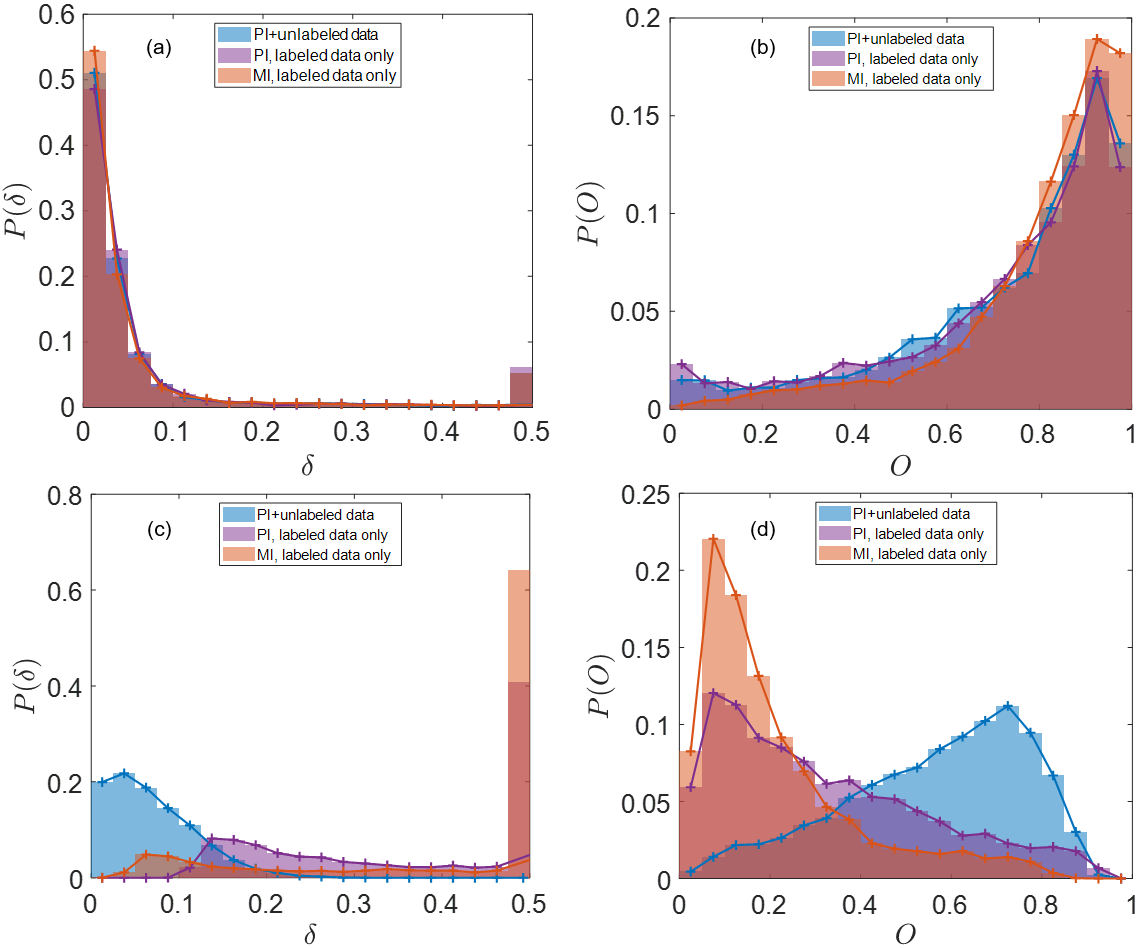}
\caption{Performance of ML models trained on plasmonic dataset in predicting the modes of plasmonic (a,b) and low-index photonic (c,d) geometries; low-index photonic configurations (without corresponding mode solutions) are used to supplement training of {\it  PI+unlabeled data} model; last bin in panels (a,c) represents data with $\delta\ge 0.5$ }.
\label{figPG}
\end{figure}

However, the models trained on high-index plasmonic or photonic datasets perform poorly when they are deployed to analyze the modes of low-index configurations and vice versa. This phenomenon is illustrated in Fig.\ref{figPG}. Once again, the performance of the models based on physics-informed training can be significantly improved by expanding their original plasmonic training set with unlabeled low-index configurations. Note that such an improvement in model generalizability does not affect model's performance on its original dataset. 

Overall, our analysis suggests that unlabeled subset promotes physics-consistency, with resulting models correctly predicting eigenvalue/eigenvector pairs of Eq.\eqref{eq:3}. However, in the absence of sufficient labeled data, the models often fail to predict the particular solution representing the largest-$k_y$ propagating mode within the spectrum, especially in composites where multiple modes with similar propagation constants are supported. Convergence of ML-based models to proper modes can be improved by either the techniques introduced in Ref.\cite{VT} (such as introducing the eigenvalue ``pull'' terms into loss function), or by expanding labeled training subset.


Along with prediction accuracy, prediction speed represents another major factor in practical applications of mode analyzers, such that optimizing the geometry of the composite to achieve the particular field distribution, mode confinement, propagation constant, etc. Here, pre-trained ML models are drastically faster than their direct Physics-solver counterparts (it takes $\sim 0.3$s to predict the properties of the high-$k_y$ modes in all 8000 elements of a given set on our desktop with Intel Core I7-10700 processor and NVidia GeForce RTX-3060 GPU), compared with $\sim 80$s it takes to run RCWA algorithm. In our tests the prediction speed was almost independent of the ML solver as well as of the complexity of the problem (see below); given previous research we expect the prediction speed to grow when the size of the dataset is significantly increased (to $\sim 10^5\ldots 10^6$ configurations). 

The time it takes to train the model depends not only on the size of the training set but also on the method used. ``Black box'' and Meaning-informed models train virtually in the same time ($\sim 200\ldots 250$s for 5000 epochs). The Physics-informed model trains substantially slower (300...600s for 5000 epochs, depending on the size of the training dataset).   

Therefore, when the number of configurations under study is small, it is beneficial to use direct solvers. Once the number of configurations reaches some critical value, ML-based tools increasingly outperform their physics-based counterparts in terms of the overall training+prediction time. For the 1D toy model of moderate complexity considered in this work, ML-learning tools ``break even'' with brute-force solvers when the number of configurations reaches $\sim 2\times 10^4$.  

\begin{figure}
\centering
\includegraphics[width=8cm]{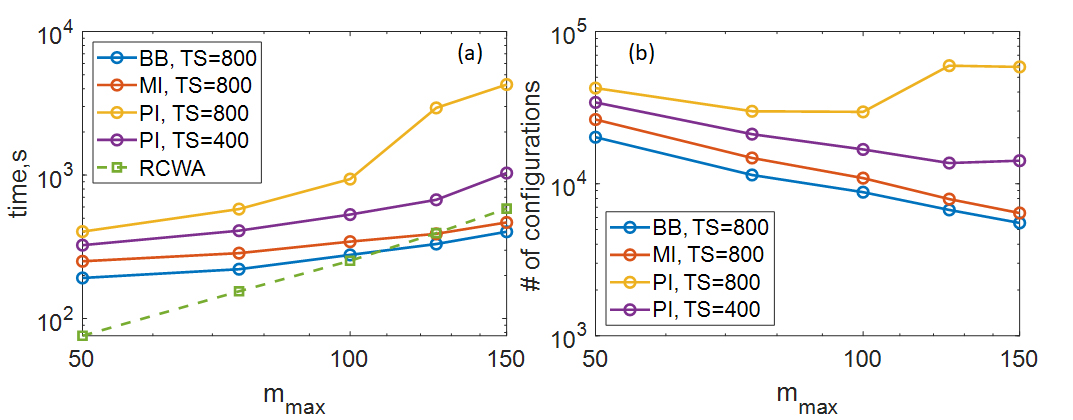}
\caption{(a) comparison of time it takes to calculate the properties of 8000 modes with RCWA and to train ML algorithms on 10\% and 5\% of these configurations (it takes 0.3s to predict 8000 modes with all ML models in our studies). (b) the number of configurations when time it takes to run RCWA calculations breaks even with the time it takes to train the ML model on a subset of the configurations and use it to predict the mode properties of the remaining configurations; in both panels TS stands for the size of the training set}.
\label{figTiming}
\end{figure}

The advantage of ML tools grows as complexity of the composite increases. For example, increasing the dimensionality of the eigenvector (by increasing the parameter $m_{\rm max}$) strongly affects the RCWA runtime. However, the time required to train black box- and meaning-informed-models grows at the slower rate, while the  time it takes the model to predict the solutions are virtually unaffected by these changes (see Fig.\ref{figTiming}). The ``breaks even points'' for BB and MI models steadity decrease with the increase of $m_{\rm max}$.

For relatively simple systems, the PI models behave similar to their MI and BB counterparts. As result, we conclude that the training time of these models is dominated by calculations of gradients and adjustments of learning parameters for ANN. For more complex systems ($m_{\rm max}\gtrsim 100$), the dependence of the training time the PG models on the complexity becomes similar to that of RCWA, reflecting the regime where training process is dominated by the calculations of Physics Loss. In our studies, PI models with large training sets are also affected by hardware constraints: the model reaches the available GPU memory when $m_{\rm max}\simeq 100$, explaining the rapid rise of the training time due to memory-swapping. 

Overall, it is seen that the ML-based solutions provide meaningful speedup for large-scale exploratory studies of optics of composites, especially when a pre-trained model is used.

\section{Conclusions}
 We developed an approach to introduce Physics-based constraints into ML algorithms for analysis of optical composites and demonstrated that Physics-informed ML models provide better generalizability and better accuracy than their ``black box'' counterparts. The utility of the Physics-informed ML has been illustrated on example of calculating the properties of the highest-index mode of the periodic multilayered composites, where pre-trained ML models offer orders of magnitude (0.3s vs 80s) speedup over conventional numerical solutions of Maxwell equations. The developed formalism is directly applicable to calculation of modes in arbitrary periodic composites and can be extended to non-periodic and guiding structures, resolving crucial computational bottlenecks in design and optimization of composite-based applications. 
 
 More importantly, the ability to introduce Physics-based constraints into ML algorithms provides the pathway to merge the benefits of powerful pattern-recognition-based learning that are inherent to ML with the benefits of analytical scientific knowledge that has been accumulated within Physics community. 
 
\subsubsection{Funding}
This research is supported by the NSF (Grants \#III-2026703, \#IIS-2026702, \#2026710).

\subsubsection{Data Availability Statement}
The codes used in the project, along with data underlying the results are available in Ref.\cite{githubPGMLvp} and may be  obtained from the authors upon reasonable request.


\appendix

\begin{figure*}[tbh]
    \centering
    \includegraphics[width=0.8\textwidth]{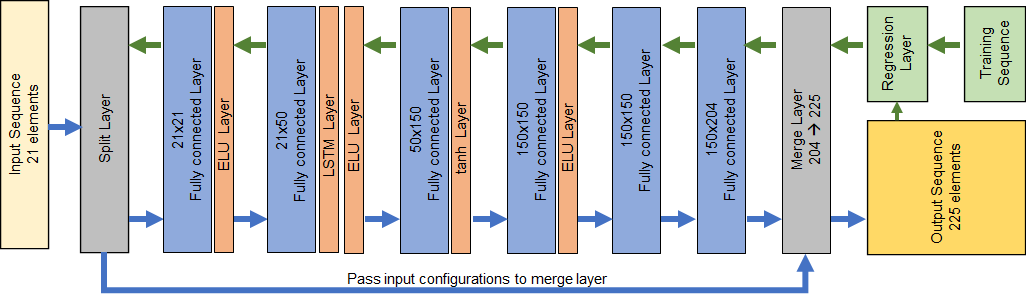}
    \caption{Schematic diagram of the Artificial Neural Network utilized in this work; blue and green arrows represent forward (prediction) and backward (training) pass, respectively. BB, MI, and PI models differ by implementation of the the loss function within the regression layer. Training sequence contain configuration $\rightarrow$ solution pairs for all models, as well as extra unlabeled configurations for PI models}
    \label{dl Network}
\end{figure*}

\section{Modal Overlap}
To assess the quality of the predicted field distribution of the electromagnetic mode provided by the machine learning (ML) algorithms, we calculate the [normalized] overlap between the predicted and the exact field distributions. Explicitly, 
\begin{equation}
    O=\frac{\left|\frac{1}{L}\int_0^L H_z^*(x) H_z^{\rm gt}(x) dx\right|}
    {\sqrt{\frac{1}{L}\int_0^L \left|H_z(x)\right|^2 dx}
    \sqrt{\frac{1}{L}\int_0^L \left|H_z^{\rm gt}(x)\right|^2 dx}},
    \label{eqOspace}
\end{equation}
where superscripts $\rm gt$ and $^*$ represent ground truth and complex conjugation, respectively. Direct substitution of  Eq.(2) from the main manuscript into Eq. \eqref{eqOspace} yields representation of the overlap integral in terms of Fourier components $h_m$ [Eq. (5) from the main manuscript]
(note that the eighenvectors produced by both RCWA and ML algorithms are normalized so that $\sum_m |h_m|^2=1$).

The value $O=1$ represents the perfect match between the predicted field profile and the ground truth (with exception of a multiplicative complex-valued constant). The values $O\ll 1$ represent poor match between the two field profiles. Fig.3 of the main manuscript illustrates this dynamics.

\section{Artificial Neural Network setup}

The Artificial Neural Networks (ANNs) used in this work contain a set of  fully-connected layers, coupled via LSTM and nonlinear activation layers, as illustrated in Fig.\ref{dl Network}. In order to facilitate physics-informed learning, we use custom layer that duplicates input sequence and carries one (unchanged) copy of this sequence towards the output layer. Our  implementation of the ANNs is available on GitHub\cite{githubPGMLvp}. The topology of the network remained the same through all studies in this work, with the only difference between the ANNs being the calculation of the loss within the regression layer (as described below). 

\section{Black-Box Models}

\subsection{Definition of Loss-function}
This baseline physics-agnostic approach relies on the default ``black box'' regression layer that implements half-mean-squared deviation between the output of the network and the ground truth. Explicitly, for every member of the training set, the loss is calculated via 
 \begin{equation}
     L_{BB}={\frac{1}{2}\sum_{m}{\left(y^{\rm gt}_{m}-y_{m}\right)^2}}
\label{eqBBloss}
 \end{equation}
 where $y_{m}$ specifies the $m^{th}$ element of the output sequence and superscript ``gt'' represents ground truth data. When the training set contains data for multiple configurations, the overall loss is calculated as a mean of the configuration-specific loss, given by the Eq.(\ref{eqBBloss}).
  
 \subsection{Performance of the Black-Box models}
\begin{figure*}[bth]
    \centering
    \includegraphics[width=0.8\textwidth]{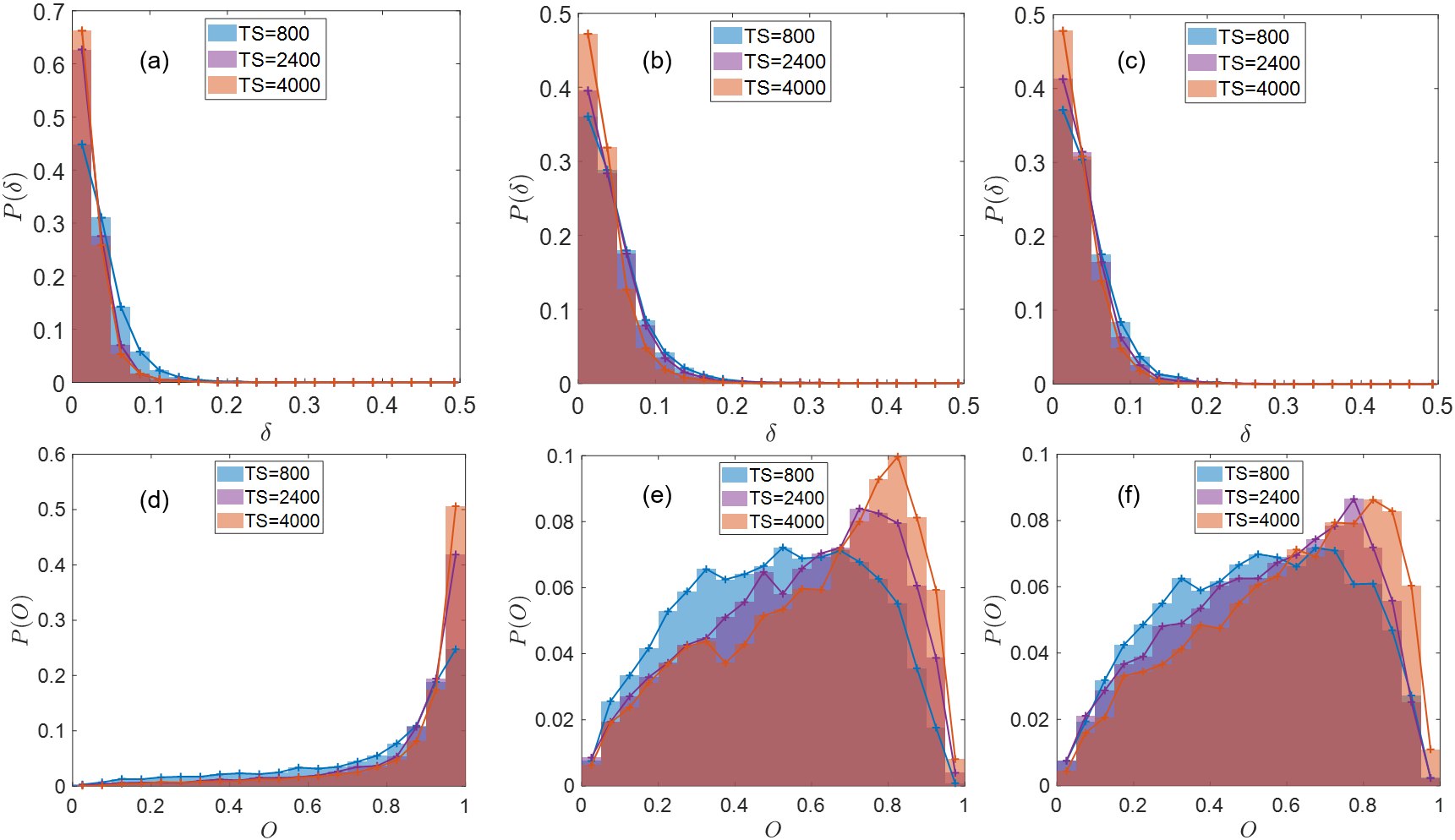}
    \caption{Performance of the Black-box model as a function of the size of the training set for composites of different complexities, corresponding to different value of $m_{max}$. All figures represent data for low-index photonic composites, with $m_{\rm max}=10$ (a,d); $m_{\rm max}=25$ (b,e), and $m_{\rm max}=50$ (c,f). The size of the training set is indicated in the legends. See manuscript for definitions of parameters $\delta$ and $O$ 
    }
     \label{BB}
\end{figure*}

Performance of any data-driven prediction depends on the size and quality of its training set, as well as on the complexity of the task at hand. To assess the relationship between these parameters we analyzed performance of the black-box models (used as baselines in our studies) as functions of both the training set and the parameter $m_{\rm max}$ that determines the number of components in the Fourier expansions used in RCWA algorithm. The results of these studies are illustrated in Fig.\ref{BB}

As expected, both reduction of the complexity of the model (representing the smaller $m_{\rm max}$) and the increase of the training size yield the models with higher accuracy. Note however that standard physics-agnostic models tend to require rather large training sets in order to generate acceptable results. Therefore, even for toy models of moderate complexity ($m_{\rm max}=50$) ``black box'' models provide only limited practical value. 

  \section{Meaning-Informed Models}
The sequence of elements $y_m$ produced by the ANN in this work contains $4m_{\rm max}+25$ elements per configuration. The first two elements of this sequence describe the propagation constant of the mode, with $k_y=\frac{\omega}{c}(y_1+y_2 i)$. The following $4m_{\rm max}+2$ elements represent, respectively, the real ($y_3 \ldots y_{2 m_{\rm max}+3}$) and imaginary ($y_{2 m_{\rm max}+4}\ldots y_{4 m_{\rm max}+4}$) parts of the coefficients $(h_{-m_{\rm max}}\ldots h_{m_{\rm max}})$ that describe the field profile of the mode. The final 21 elements of the $y$ sequence describe the parameter $\theta= y_{4 m_{\rm max}+5}$ and real ($y_{4 m_{\rm max}+6}\ldots y_{4 m_{\rm max}+15}$) and imaginary ($y_{4 m_{\rm max}+16}\ldots y_{4 m_{\rm max}+25}$) parts of $\epsilon(x)$ representing each of the ten ``pixels'' of the layered structure. 
 
 Note that while configuration array remains part of the output sequence, its elements are automatically appended to the predictions of the ANN via split/merge layer pair. The learning parameters within the ANN are only trained to predict the elements ($y_1\ldots y_{4 m_{\rm max}+4}$) of the output. 
 
  \subsection{Definition of Loss-function}

  The physical meaning of the different elements of the output sequence has been incorporated into the loss function of the ANN by creating a custom regression layer where the loss of the particular configuration is represented as: 
  \begin{eqnarray*}
\\
L_{MI}=w_{\delta}\delta +w_O(1-O)^2.
  \end{eqnarray*}
 with parameters $w_\delta$ and $w_O$ representing the relative weights of the loss associated with accuracy of prediction of the propagation constant and the field profile, respectively (see Eq. (5) in the main manuscript). Our analysis suggests that the quality of the predictions as well as training dynamics only weakly depend on the numerical values of the weight factors. The final analysis shown in this work represents data produced with $w_\delta=1$ and $w_O=5$.

\begin{figure*}[tbh]
    \centering
    \includegraphics[width=0.8\textwidth]{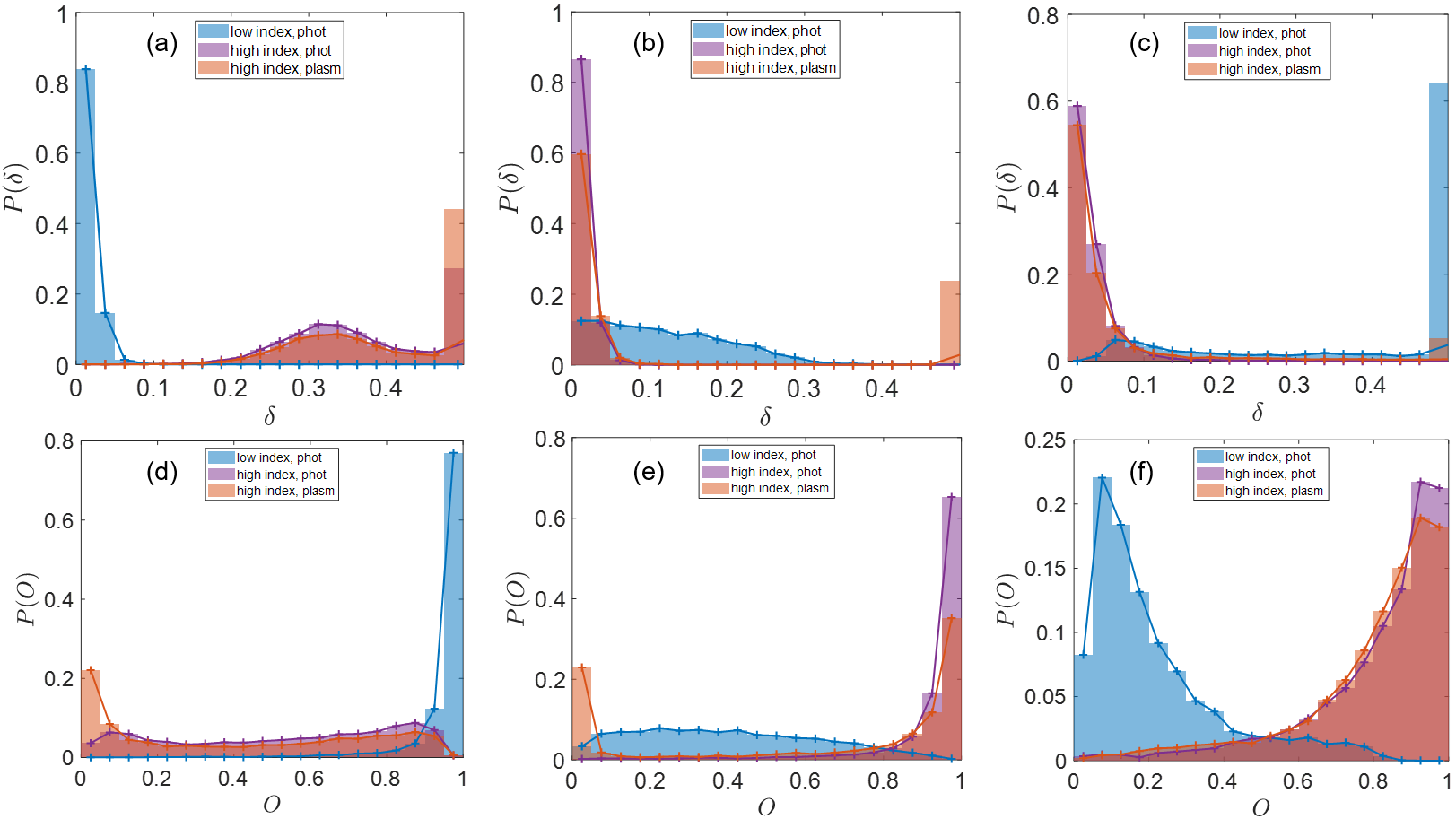}
    \caption{ Performance of Meaning-Informed ANNs when predicting properties of highest-index mode in low-index photonic, high-index photonic, and high-index plasmonic datasets (legends). Panels (a,d), (b,e), and (c,f) represent results of MI models trained on a portion of low-index photonic , high index photonic , and high index plasmonic datasets, respectively. In all cases $m_{max}=50$; $P(\delta)$ and $P(O)$ are the probability distribution with respect to the deviation of propagating constant $\delta$ and modal overlap $O$ respectively.}
    \label{MG}
\end{figure*}
  
  \subsection{Performance of Meaning-Informed models}
Fig.\ref{MG} illustrates the performance of the Meaning-informed models trained on $800\ldots 2400$ configurations. It is seen the performance of the MI models is significantly better than the performance of the comparably trained BB model. Notably, the plasmonic dataset, due to the diversity of its configurations, and due to extremely high-index SPP modes present in these configurations, is the most challenging one for ML to predict. 

It is also important to notice that ML models perform well only on the dataset that they have been exposed to during the training. For example, the ANN trained on low-index data fails to predict high-index data and vice versa (as mentioned in the main manuscript, high index plasmonic dataset contains significant number of elements from high-index photonic dataset).

  \section{Physics-Informed Models}
  \subsection{Definition of Loss-function}
  
  In addition to meaning-informed loss, Physics-Informed models incorporate learning objectives inspired by analytical equations underlying the solutions of Maxwell equations. Explicitly, we introduce Physics Loss (PL) that enforces Eq.(3) of the main manuscript via
  \begin{equation}
     L_P=\frac{\sum_{m}\left|\sum_{j}\hat{A}_{m j} h_j-k_y^2 h_m\right|^2}
     {\sum_m\left|\sum_j\hat{A}_{m j} h_j\right|^2}.
     \label{eqPLoss}
  \end{equation}
  
  Importantly, PL defined by Eq.(\ref{eqPLoss}), can be used to assess physics-consistency of the solutions produced by the ANN even in absence of the ground truth (on unlabeled datasets). Therefore, PL can be utilized to expand the training set beyond the constraints of pre-calculated data. However, since Eq.(\ref{eqPLoss}) is  satisfied for any electromagnetic mode supported by the periodic composite, this equation by itself cannot ensure that the solution corresponds to the mode with largest propagating constant. Similar situation appeared in the previous work where ANN was trained directly on matrices $\hat{A}$\cite{VT}, where it was shown that the loss landscape formed by the combination of $L_{P}$ and $L_{BB}$ contains multiple local minima representing these ``side'' solutions of Maxwell equations. 
  
  To force the ANN towards the highest-propagation constant mode, we followed the approach of Ref.\cite{VT} and introduced spectrum loss $L_S$ that is used at the initial training stage, 
  \begin{equation*}
   L_S=e^{-{\rm Re}(k_{y}c/\omega)}+{\rm Im}(k_y c/\omega)^4.
  \end{equation*}
  
  The full loss of PI model combines its MI counterpart, as well as {\em dynamically-weighed} contributions of the Physics- and Spectrum losses,
   \begin{eqnarray*}
  L_{PI}=L_{MI}+w_P(t)L_P+w_S(t) L_S, 
  \end{eqnarray*}
  with  $w_S(t)=w_S^0\exp{(-t/t_S^0)}$, $w_P(t)=w_P^0[1+\exp{(-(t-t_P^i)/t_P^0)]^{-1}}$, and $t$ being the training epoch index, enforcing the training process that is affected by the spectrum loss only at the beginning stages of the training and by PL at the later stages. The parameters that define the dynamics of the training were set to $t_S^0=100, t_P^i=150, t_P^0=300$, while the relative weights for Spectrum and Physics Loss were fixed at $w_S^0=75,w_P^0=1$ for unlabeled training configurations and $w_S^0=0,w_P^0=0.5$ for labeled configurations. Similar to BB and MI models, the total loss for the training set is computed as arithmetic mean of the configuration-specific loss.

\begin{figure}[tbh]
    \centering
    \includegraphics[width=8cm]{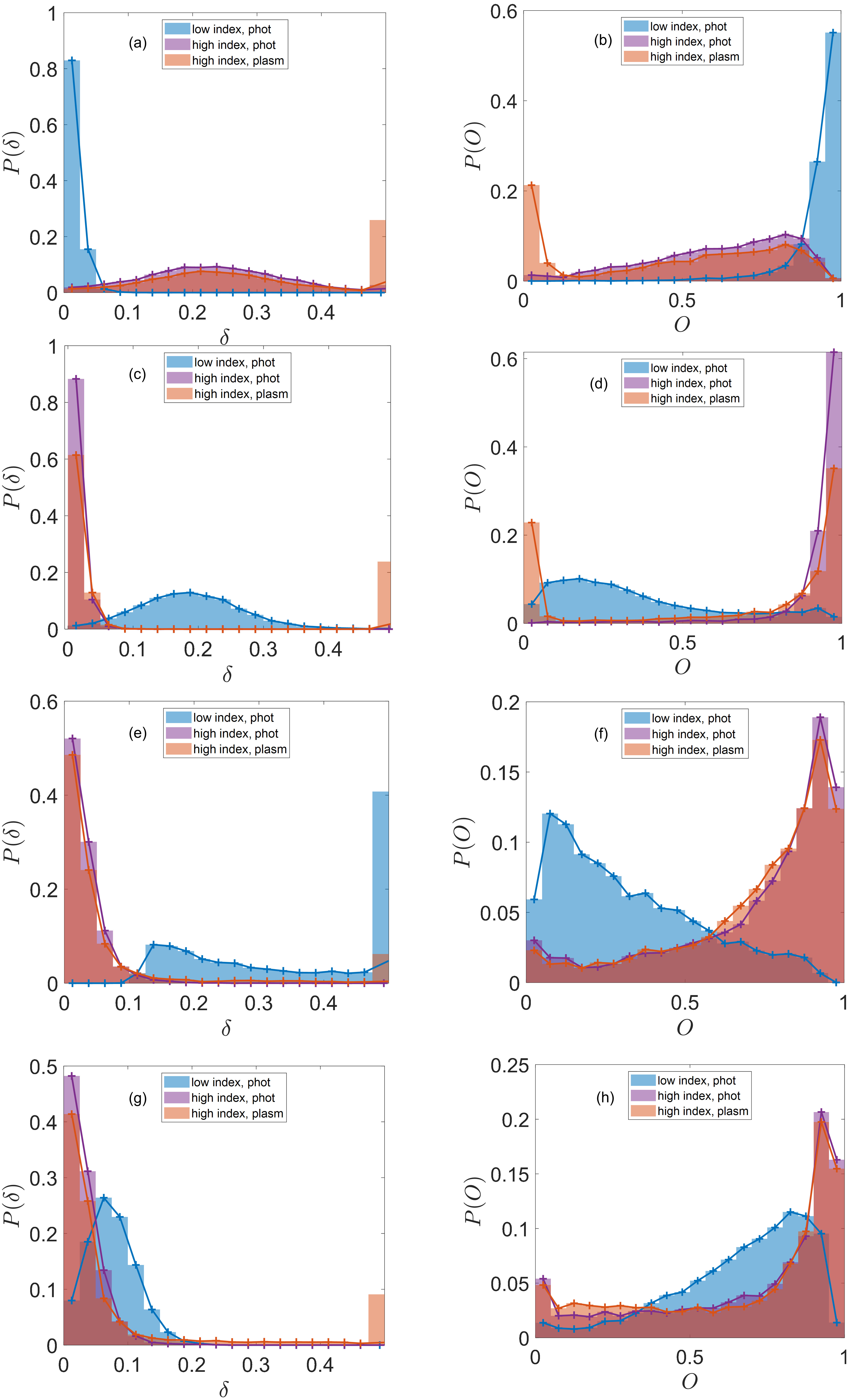}
     \caption{
      Performance of Physics-Informed ANNs when predicting properties of highest-index mode in low-index photonic, high-index photonic, and high-index plasmonic datasets (legends). Panels (a\ldots f) summarize the outcomes of PI models trained on fully-labeled portions of low-index photonic (a,b), high index photonic (c,d), and high index plasmonic (e,f) datasets, respectively. Panels (g,h) represent the result of PI model trained on a portion of high-index plasmonic data with added (unlabeled) configurations from low-index photonic structures. In all cases $m_{max}=50$; $P(\delta)$ and $P(O)$ are the probability distribution with respect to the deviation of propagating constant $\delta$ and modal overlap $O$ respectively.
     }
     \label{PG}
\end{figure}

  
  
\subsection{Performance and Generalizability of the Physics-Informed Deep Learning ANN-Algorithm}
In general, performance of the PI models trained exclusively on known solutions (labeled data) closely matched the performance of their MI counterparts. Summary of these results is shown in Fig.\ref{PG}(a\ldots f). It is seen that, similar to MI models, the PI models trained exclusively on labeled data do not generalize well beyond the configurations seen by the model in the training set. 

However, as discussed in the main text, the training set of PI models can be expanded by providing configurations with no known solutions as unlabeled data. As seen in Fig.\ref{PG}(g,h), such procedure  significantly improves stability and generalizability of the resulting model.

\begin{figure}[bh]
    \centering
    \includegraphics[width=8cm]{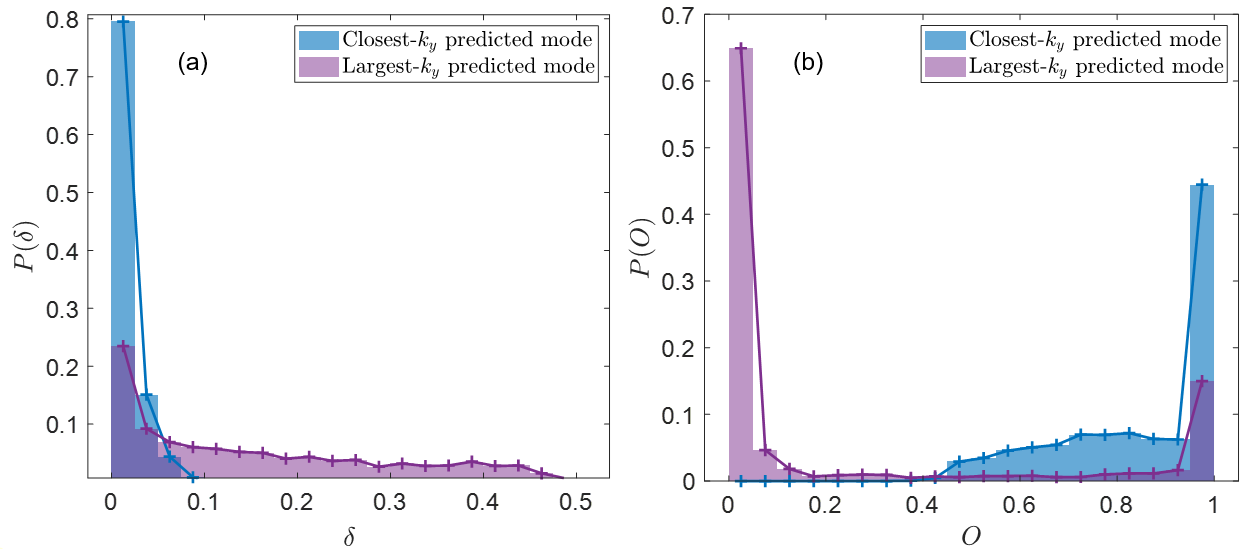}
    \caption{Performance of the PI model, trained exclusively on unlabeled data (representing subset of configurations from low index photonic dataset); the curves represent distriution of propagation constant deviations (a) and overlap integrals (b) for ANN predictions, as compared with ground truth for highest-propagating constant mode supported by the composite and for the mode supported by the same composite that has the spatial profile closest to ANN predictions}
    \label{unLabeled data only}
\end{figure}

Finally, we note that PI models can be trained exclusively on unlabeled data. Fig.\ref{unLabeled data only} illustrates a result of such a procedure. It is clear that the model, as trained, does not provide the solution for the highest-index eigen-mode of the composite. To take a closer look at the ANN predictions we calculated the full spectrum of the eigensolutions for each configuration. We then identified the mode that is closest to the one predicted by the ANN, and calculated the parameters $\delta$ and $O$ for that particular mode. The distribution of these parameters are shown in Fig.\ref{unLabeled data only}. It is seen that the ML model does converge to a valid propagation constant/field distribution pair, it just fails to
identify the highest-eigenvalue mode. 

Our analysis indicates that while the addition of spectrum loss
somewhat mitigates this problem, the relatively dense spectrum of the propagation constants, and
therefore a complex loss function topology, makes it extremely challenging to guide the model
towards the right (and correct) mode without extra constraints that are provided by labeled data
in models shown in Fig.\ref{PG}.

 \end{document}